\documentclass[journal,pdf]{IEEEtran}

\usepackage{cite}

\ifCLASSINFOpdf
  \usepackage[pdftex]{graphicx}
\else
\fi

\usepackage{amsmath}
\usepackage{amssymb}

\usepackage{url}
\begin{document}
\title{Deep Convolutional Neural Networks and Data Augmentation for Environmental Sound Classification}

\author{Justin~Salamon 
        and~Juan~Pablo~Bello% <-this % stops a space
\thanks{J.~Salamon (justin.salamon@nyu.edu) is with the Music and Audio Research Laboratory (MARL) and the Center for Urban Science and Progress (CUSP) at New York University, USA. J.~P.~Bello (jpbello@nyu.edu) is with the Music and Audio Research Laboratory at New York University, USA.}
}

\markboth{IEEE Signal Processing Letters, accepted November 2016}%
{Salamon \& Bello: Deep Convolutional Neural Networks and Data Augmentation for Environmental Sound Classification}

\maketitle

\begin{abstract}
The ability of deep convolutional neural networks (CNN) to learn discriminative spectro-temporal patterns makes them well suited to environmental sound classification. However, the relative scarcity of labeled data has impeded the exploitation of this family of high-capacity models.
This study has two primary contributions: first, we propose a deep convolutional neural network architecture 
for environmental sound classification. Second, we propose the use of audio data augmentation for overcoming the problem of data scarcity and explore the influence of different augmentations on the performance of the proposed CNN architecture.
Combined with data augmentation, the proposed model produces state-of-the-art results for environmental sound classification. We show that the improved performance stems from the combination of a deep, high-capacity model and an augmented training set: this combination outperforms both the proposed CNN without augmentation and a ``shallow'' dictionary learning model with augmentation. Finally, we examine the influence of each augmentation on the model's classification accuracy for each class, and observe that the accuracy for each class is influenced differently by each augmentation, suggesting that the performance of the model could be improved further by applying class-conditional data augmentation.
\end{abstract}


\ifCLASSOPTIONpeerreview
\begin{center} \bfseries EDICS Category: 3-BBND \end{center}
\fi
\IEEEpeerreviewmaketitle

\section{Introduction}
\label{sec:intro}

\IEEEPARstart{T}{he} problem of automatic environmental sound classification has received increasing attention from the research community in recent years.
Its applications range from context aware computing \cite{chu:EnvironmentalSoundRec:TASLP:09} and surveillance \cite{radhakrishnan:AudioSurveillance:WASPAA:05} to noise mitigation enabled by smart acoustic sensor networks \cite{Mydlarz:AcousticMonitoring:AA:16}.

To date, a variety of signal processing and machine learning techniques have been applied to the problem, including matrix factorization \cite{Mesaros:SEDNMF:ICASSP:15,Benetos:AED:ICASSP:16,Bisot:AcousticSceneNMF:ICASSP:16},  dictionary learning \cite{Salamon:UnsupervisedUrban:ICASSP:15,Salamon:ScattteringUrban:EUSIPCO:15}, wavelet filterbanks \cite{Geiger:AEDGabor:EUSIPCO:15,Salamon:ScattteringUrban:EUSIPCO:15} and most recently deep neural networks \cite{Cakir:SEDDNN:IJCNN:15,Piczak:EnvSoundCNN:MLSP:15}. See \cite{giannoulis:AASPChallenge:WASPAA:13,Stowell:AEDreview:TMM:15,Sigtia:AEDPerformanceVsComp:TASLP:16} for further reviews of existing approaches.
In particular, deep convolutional neural networks (CNN) \cite{Lecun:CNN:IEEE:98} are, in principle, very well suited to the problem of environmental sound classification: first, they are capable of capturing energy modulation patterns across time and frequency when applied to spectrogram-like inputs, which has been shown to be an important trait for distinguishing between different, often noise-like, sounds such as engines and jackhammers \cite{Salamon:ScattteringUrban:EUSIPCO:15}. Second, by using convolutional kernels (filters) with a small receptive field, the network should, in principle, be able to successfully learn and later identify spectro-temporal patterns that are representative of different sound classes even if part of the sound is masked (in time/frequency) by other sources (noise), which is where traditional audio features such as Mel-Frequency Cepstral Coefficients (MFCC) fail \cite{cotton:SpectroTemporal:WASPAA:11}. Yet the application of CNNs to environmental sound classification has been limited to date. For instance, the CNN proposed in \cite{Piczak:EnvSoundCNN:MLSP:15} obtained comparable results to those yielded by a dictionary learning approach \cite{Salamon:UnsupervisedUrban:ICASSP:15} (which can be considered an instance of ``shallow'' feature learning), but did not improve upon it.

Deep neural networks, which have a high model capacity, are particularly dependent on the availability of large quantities of training data in order to learn a non-linear function from input to output that generalizes well and yields high classification accuracy on unseen data.
A possible explanation for the limited exploration of CNNs and the difficulty to improve on simpler models is the relative scarcity of labeled data for environmental sound classification. While several new datasets have been released in recent years (e.g., \cite{Salamon:UrbanSound:ACMMM:14,Piczak:ESCdataset:ACMMM:15,Mesaros:TUT16dataset:Zenodo:16}), they are still considerably smaller than the datasets available for research on, for example, image classification \cite{Krizhevsky:Imagenet:NIPS:2012}. 

An elegant solution to this problem is \emph{data augmentation}, that is, the application of one or more deformations to a collection of annotated training samples which result in new, additional training data \cite{Simard:CNNBestPractices:ICDAR:03,Krizhevsky:Imagenet:NIPS:2012,McFee:Augmentation:ISMIR:15}. A key concept of data augmentation is that the deformations applied to the labeled data do not change the semantic meaning of the labels. Taking an example from computer vision, a rotated, translated, mirrored or scaled image of a car would still be a coherent image of a car, and thus it is possible to apply these deformations to produce additional training data while maintaining the semantic validity of the label. By training the network on the additional deformed data, the hope is that the network becomes invariant to these deformations and generalizes better to unseen data. Semantics-preserving deformations have also been proposed for the audio domain, and have been shown to increase model accuracy for music classification tasks \cite{McFee:Augmentation:ISMIR:15}. However, in the case of environmental sound classification the application of data augmentation has been relatively limited (e.g., \cite{Piczak:EnvSoundCNN:MLSP:15,Parascandolo:RNNSED:ICASSP:16}), with the author of \cite{Piczak:EnvSoundCNN:MLSP:15} (which used random combinations of\linebreak time shifting, pitch shifting and time stretching for data augmentation) reporting that ``simple augmentation techniques proved to be unsatisfactory for the UrbanSound8K dataset given the considerable increase in training time they generated and negligible impact on model accuracy''.

In this paper we present a deep convolutional neural network architecture 
with localized (small) kernels
for environmental sound classification. Furthermore, we propose the use of data augmentation to overcome the problem of data scarcity and explore different types of audio deformations and their influence on the model's performance. We show that the proposed CNN architecture, in combination with audio data augmentation, yields state-of-the-art performance for environmental sound classification.

\section{Method}
\label{sec:method}

\subsection{Deep Convolutional Neural Network}

The deep convolutional neural network (CNN) architecture proposed in this study 
is comprised of 3 convolutional layers interleaved with 2 pooling operations, followed by 2 fully connected (dense) layers.
Similar to previously proposed feature learning approaches applied to environmental sound classification (e.g., \cite{Salamon:UnsupervisedUrban:ICASSP:15}), the input to the network consists of time-frequency patches (TF-patches) taken from the log-scaled mel-spectrogram representation of the audio signal. Specifically, we use Essentia \cite{bogdanov:Essentia:ISMIR13} to extract log-scaled mel-spectrograms with 128 components (bands) covering the audible frequency range (0-22050 Hz), using a window size of 23 ms (1024 samples at 44.1 kHz) and a hop size of the same duration. Since the excerpts in our evaluation dataset (described below) are of varying duration (up to 4 s), we fix the size of the input TF-patch $X$ to 3 seconds (128 frames), i.e.~$X\in\mathbb{R}^{128\times 128}$. TF-patches are extracted randomly (in time) from the full log-mel-spectrogram of each audio excerpt during training as described further down.

Given our input $X$, the network is trained to learn the parameters $\Theta$ of a composite nonlinear function $\mathcal{F}(\cdot|\Theta)$ which maps $X$ to the output (prediction) $Z$:
\begin{equation}
Z = \mathcal{F}(X|\Theta) = f_L(\cdots f_2(f_1(X|\theta_1)|\theta_2)|\theta_L),
\end{equation}
where each operation $f_\ell(\cdot|\theta_\ell)$ is referred to as a \emph{layer} of the network, with $L=5$ layers in our proposed architecture. The first three layers, $\ell\in\{1,2,3\}$, are convolutional, expressed as:
\begin{equation}
Z_\ell = f_\ell(X_\ell|\theta_\ell) = h(W \ast X_\ell + b),\quad\theta{}_l=\lbrack W,b\rbrack
\end{equation}
where $X_\ell$ is a 3-dimensional input tensor consisting of $N$ \emph{feature maps}, $W$ is a collection of $M$ 3-dimensional kernels (also referred to as filters), $\ast$ represents a valid convolution, $b$ is a vector bias term, and $h(\cdot)$ is a point-wise activation function. Thus, the shapes of $X_\ell$, $W$, and $Z_\ell$ are $(N, d_0, d_1)$, $(M, N, m_0, m_1)$ and $(M, d_0-m_0+1, d_1-m_1+1)$ respectively. Note that for the first layer of our network $d_0 = d_1 = 128$, i.e., the dimensions of the input TF-patch. 
We apply strided
max-pooling after the first two convolutional layers $\ell\in\{1,2\}$
using a stride size equal to the pooling dimensions (provided below),
which reduces the dimensions of the output feature maps and consequently speeds up training and builds some scale invariance into the network. The final two layers, $\ell\in\{4,5\}$, are fully-connected (dense) and consist of a matrix product rather than a convolution:
\begin{equation}
Z_\ell =  f_\ell(X_\ell|\theta_\ell) = h(WX_\ell+b),\quad \theta_\ell=\lbrack W,b\rbrack
\end{equation}
where $X_\ell$ is flattened to a column vector of length $N$, $W$ has shape $(M, N)$, $b$ is a vector of length $M$ and $h(\cdot)$ is a point-wise activation function.

The proposed CNN architecture is parameterized as follows:
\begin{itemize}
\item $\ell_1$: 24 filters with a receptive field of (5,5), i.e., $W$ has the shape (24,1,5,5). This is followed by (4,2) strided max-pooling over the last two dimensions (time and frequency respectively) 
and a rectified linear unit (ReLU) activation function $h(x) = \max(x,0)$.
\item $\ell_2$: 48 filters with a receptive field of (5,5), i.e., $W$ has the shape (48, 24, 5, 5). Like $\ell_1$, this is followed by (4,2) strided max-pooling and a ReLU activation function.
\item $\ell_3$: 48 filters with a receptive field of (5,5), i.e., $W$ has the shape (48, 48, 5, 5). This is followed by a ReLU activation function (no pooling).
\item $\ell_4$: 64 hidden units, i.e., $W$ has the shape (2400, 64), followed by a ReLU activation function.
\item $\ell_5$: 10 output units, i.e., $W$ has the shape (64,10), followed by a softmax activation function.
\end{itemize}
Note that our use of a small receptive field $(5,5)$ in $\ell_1$ compared to the input dimensions $(128,128)$ is designed to allow the network to learn small, localized patterns that can be fused at subsequent layers to gather evidence in support of larger ``time-frequency signatures'' that are indicative of the presence/absence of different sound classes, even when there is spectro-temporal masking by interfering sources.

For training, the model optimizes cross-entropy loss via mini-batch stochastic gradient descent \cite{Bottou:SGD:COMPSTAT:10}. Each batch consists of 100 TF-patches randomly selected from the training data (without repetition). Each 3 s TF-patch is taken from a random position in time from the full log-mel-spectrogram representation of each training sample. We use a constant learning rate of 0.01. Dropout \cite{Srivastava:Dropout:JMLR:14} is applied to the input of the last two layers, $\ell \in \{4,5\}$, with probability 0.5. L2-regularization is applied to the weights of the last two layers with a penalty factor of 0.001. The model is trained for 50 epochs and is checkpointed after each epoch, during which it is trained on random minibatches until 1/8 of all training data is exhausted (where by training data we mean all the TF-patches extracted from every training sample starting at all possible frame indices).
A validation set is used to identify the parameter setting (epoch) achieving the highest classification accuracy, where prediction is performed by slicing the test sample into overlapping TF-patches (1-frame hop), making a prediction for each TF-patch and finally choosing the sample-level prediction as the class with the highest mean ouptut activation over all frames.
The CNN is implemented in Python with Lasagne \cite{Dieleman:Lasagne:ZENODO:15}, and we used Pescador \cite{McFee:Pescador:ZENODO:15} to manage and multiplex data streams during training.

\subsection{Data Augmentation}
\label{sec:method:augmentation}

We experiment with 4 different audio data augmentations (deformations), resulting in 5 augmentation sets, as detailed below. Each deformation is applied directly to the audio signal prior to converting it into the input representation used to train  the network (log-mel-spectrogram). Note that for each augmentation it is important that we choose the deformation parameters such that the semantic validity of the label is maintained. The deformations and resulting augmentation sets are described below:
\begin{itemize}
\item \textbf{Time Stretching (TS):} slow down or speed up the audio sample (while keeping the pitch unchanged). Each sample was time stretched by 4 factors: $\{0.81, 0.93, 1.07, 1.23\}$.
\item \textbf{Pitch Shifting (PS1):} raise or lower the pitch of the audio sample (while keeping the duration unchanged). Each sample was pitch shifted by 4 values (in semitones): $\{-2, -1, 1, 2\}$.
\item \textbf{Pitch Shifting (PS2):} since our initial experiments indicated that pitch shifting was a particularly beneficial augmentation, we decided to create a second augmentation set. This time each sample was pitch shifted by 4 larger values (in semitones): $\{-3.5, -2.5, 2.5, 3.5\}$.
\item \textbf{Dynamic Range Compression (DRC):} compress the dynamic range of the sample using 4 parameterizations, 3 taken from the Dolby E standard \cite{Dolby:DolbyEstandard:MISC:02} and 1 (radio) from the icecast online radio streaming server \cite{Icecast:MudaDRCSettings:MISC:16}: \{music standard, film standard, speech, radio\}.
\item \textbf{Background Noise (BG):} mix the sample with another recording containing background sounds from different types of acoustic scenes. Each sample was mixed with 4 acoustic scenes: \{street-workers, street-traffic, street-people, park\}\footnote{We ensured these scenes did not contain any of the target sound classes.}. Each mix $z$ was generated using $z=(1-w)\cdot{}x + w\cdot{}y$ where $x$ is the audio signal of the original sample, $y$ is the signal of the background scene, and $w$ is a weighting parameter that was chosen randomly for each mix from a uniform distribution in the range $\lbrack0.1,0.5\rbrack$.
\end{itemize}

The augmentations were applied using the MUDA library \cite{McFee:Augmentation:ISMIR:15}, to which the reader is referred for further details about the implementation of each deformation. MUDA takes an audio file and corresponding annotation file in JAMS format \cite{Humphrey:JAMS:ISMIR:14,McFee:JAMS:TECH:15}, and outputs the deformed audio together with an enhanced JAMS file containing all the parameters used for the deformation. We have ported the original annotations provided with the  dataset used for evaluation in this study (see below) into JAMS files and made them available online along with the post-deformation JAMS files.\footnote{\url{https://github.com/justinsalamon/UrbanSound8K-JAMS}}

\subsection{Evaluation}
\label{sec:method:evaluation}

To evaluate the proposed CNN architecture and the influence of the different augmentation sets we use the UrbanSound8K dataset \cite{Salamon:UrbanSound:ACMMM:14}. The dataset is comprised of 8732 sound clips of up to 4 s in duration taken from field recordings. The clips span 10 environmental sound classes: air conditioner, car horn, children playing, dog bark, drilling, engine idling, gun shot, jackhammer, siren and street music. By using this dataset we can compare the results of this study to previously published approaches that were evaluated on the same data, including the dictionary learning approach proposed in \cite{Salamon:UnsupervisedUrban:ICASSP:15} (spherical k-means, henceforth SKM) and the CNN proposed in \cite{Piczak:EnvSoundCNN:MLSP:15} (PiczakCNN) which has a different architecture to ours and did not employ augmentation during training.
PiczakCNN has 2 convolutional layers followed by 3 dense layers, the filters of the first layer are ``tall'' and span almost the entire frequency dimension of the input, and the network operates on 2 input channels: log mel-spectra and their deltas.

The proposed approach and those used for comparison in this study are evaluated in terms of classification accuracy. The dataset comes sorted into 10 stratified folds, and all models were evaluated using 10-fold cross validation, where we report the results as a box plot generated from the accuracy scores of the 10 folds. For training the proposed CNN architecture we use 1 of the 9 training folds in each split as a validation set for identifying the training epoch that yields the best model parameters when training with the remaining 8 folds.

\section{Results}
\label{sec:results}

\begin{figure}[!t]
\centering
\includegraphics[width=1\columnwidth,clip=true,trim=0.4cm 0.3cm 0.8cm 0.6cm ]{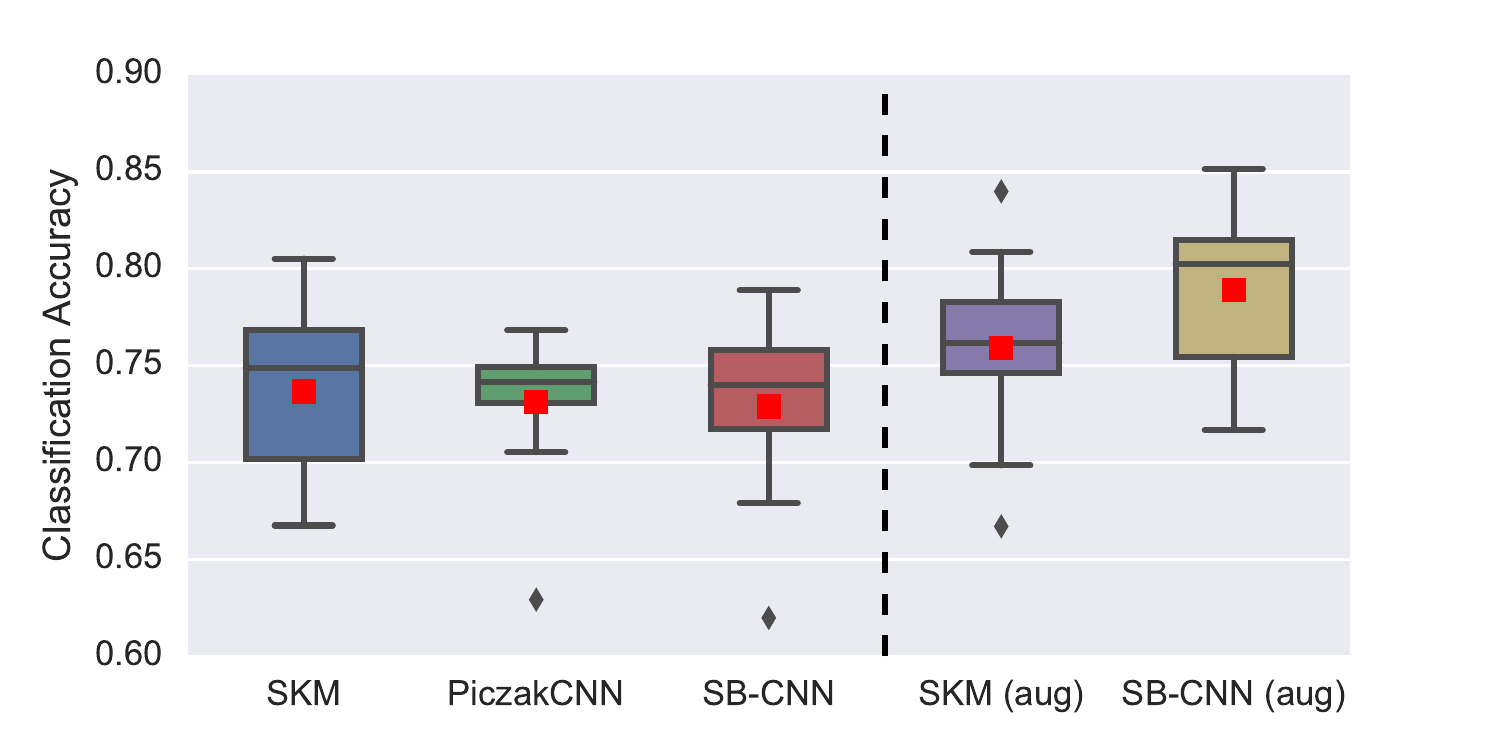}
\caption{Left of the dashed line: classification accuracy without augmentation -- dictionary learning (SKM \cite{Salamon:UnsupervisedUrban:ICASSP:15}), Piczak's CNN (PiczakCNN \cite{Piczak:EnvSoundCNN:MLSP:15}) and the proposed model (SB-CNN). Right of the dashed line: classification accuracy for SKM and SB-CNN with augmentation.}
\label{fig:accuracybox}
\end{figure}

\begin{figure}[!t]
\centering
\includegraphics[width=1\columnwidth,clip=true,trim=1.8cm 1.0cm 1.8cm 1.2cm]{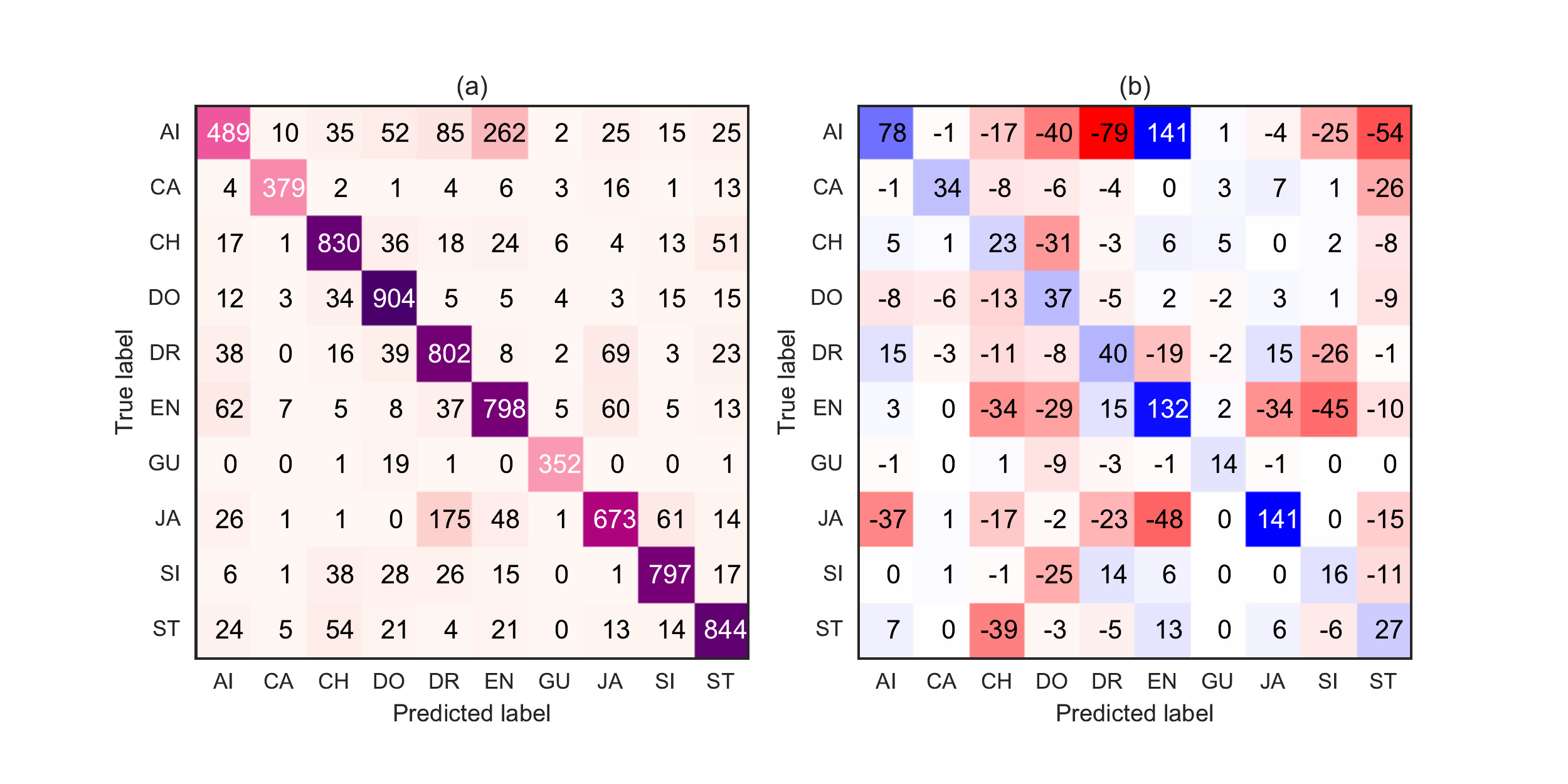}
\caption{(a) Confusion matrix for the proposed SB-CNN model with augmentation.
(b) Difference between the confusion matrices yielded by SB-CNN with and without augmentation: negative values (red) off the diagonal mean the confusion is reduced with augmentation, positive values (blue) off the diagonal mean the confusion is increased with augmentation. The positive values (blue) along the diagonal indicate that overall the classification accuracy is improved for all classes with augmentation.}
\label{fig:confusion}
\end{figure}

\begin{figure}[!t]
\centering
\includegraphics[width=0.97\columnwidth,clip=true,trim=0 0.3cm 0 0.4cm]{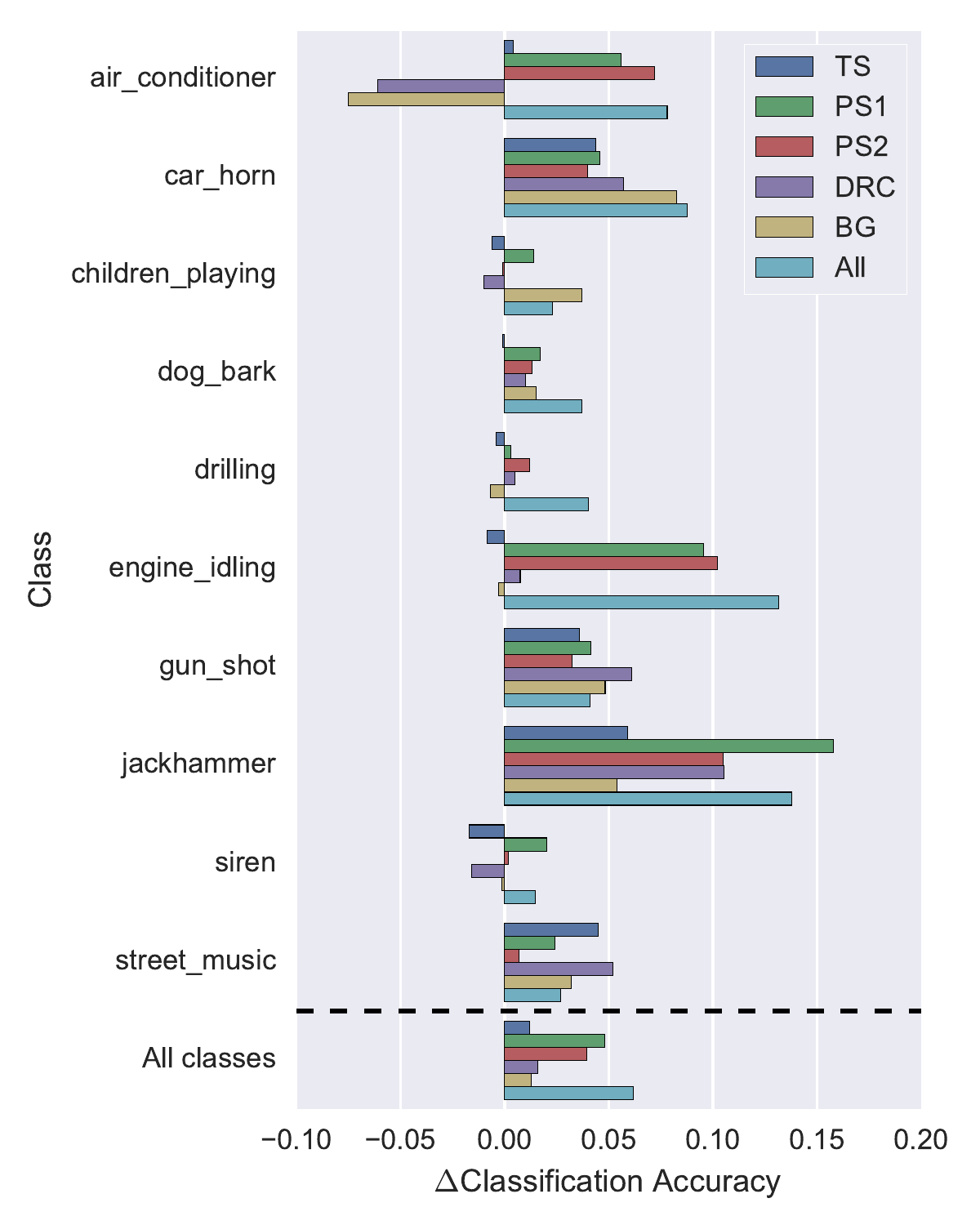}
\caption{Difference in classification accuracy for each class as a function of the augmentation applied: Time Shift (TS), Pitch Shift (PS1 and PS2), Dynamic Range Compression (DRC), Background Noise (BG) and all combined (All).}
\label{fig:accuracydelta}
\end{figure}

The classification accuracy of the proposed CNN model (SB-CNN) is presented in Figure \ref{fig:accuracybox}. To the left of the dashed line we present the performance of the proposed model on the original datast without augmentation. For comparison, we also provide the accuracy obtained on the same dataset by the dictionary learning approach proposed in \cite{Salamon:UnsupervisedUrban:ICASSP:15} (SKM, using the best parameterization identified by the authors in that study) and the CNN proposed by Piczak \cite{Piczak:EnvSoundCNN:MLSP:15} (PiczakCNN, using the best performing model variant (LP) proposed by the author). 
To the right of the dashed line we provide the performance of the SKM model and the proposed SB-CNN once again, this time when using the augmented dataset (all augmentations described in Section \ref{sec:method:augmentation} combined) for training.

We see that the proposed SB-CNN performs comparably to SKM and PiczakCNN when training on the original dataset without augmentation (mean accuracy of 0.74, 0.73 and 0.73 for SKM, PiczakCNN and SB-CNN, respectively). The original dataset is not large/varied enough for the convolutional model to outperform the ``shallow'' SKM approach. However, once we increase the size/variance in the dataset by means of the proposed augmentations, the performance of the proposed model increases significantly, yielding a mean accuracy of 0.79. 
The corresponding per-class accuracies (with respect to the list of classes provided in Section 
\ref{sec:method:evaluation}) are 0.49, 0.90, 0.83, 0.90, 0.80, 0.80, 0.94, 0.68, 0.85, 0.84.
Importantly, we note that while the proposed approach performs comparably to the ``shallow'' SKM learning approach on the original dataset, it significantly outperforms it ($p = 0.0003$ according to a paired two-sided t-test) using the augmented training set. 
Furthermore, increasing the capacity of the SKM model (by increasing the dictionary size from $k=2000$ to $k=4000$) did not yield any further improvement in classification accuracy.
This indicates that the superior performance of the proposed SB-CNN is not only due to the augmented training set, but rather thanks to the combination of an augmented training set with the increased capacity and representational power of the deep learning model. 

In Figure \ref{fig:confusion}(a) we provide the confusion matrix yielded by the proposed SB-CNN model using the augmented training set, and in Figure \ref{fig:confusion}(b) we provide the difference between the confusion matrices yielded by the proposed model with and without augmentation. 
From the latter we see that 
overall the classification accuracy is improved for all classes with augmentation. 
However, we observe that augmentation can also have a detrimental effect on the confusion between specific pairs of classes. For instance, we note that while the confusion between the air conditioner and drilling classes is reduced with augmentation, the confusion between the air conditioner and the engine idling classes is increased.

To gain further insight into the influence of each augmentation set on the performance of the proposed model for each sound class, in Figure \ref{fig:accuracydelta} we present the difference in classification accuracy (the delta) when adding each augmentation set compared to using only the original training set, broken down by sound class. At the bottom of the plot we provide the delta scores for all classes combined. We see that most classes are affected positively by most augmentation types, but there are some clear exceptions. In particular, the air conditioner class is negatively affected by the DRC and BG augmentations. Given that this sound class is characterized by a continuous ``hum'' sound, often in the background, it makes sense that the addition of background noise that can mask the presence of this class will deteriorate the performance of the model. 
In general, the pitch augmentations have the greatest positive impact on performance, and are the only augmentation sets that do not have a negative impact on any of the classes. Only half of the classes benefit from applying all augmentations combined more than they would from the application of a subset of augmentations. This suggests that the performance of the model could be improved further by the application of class-conditional augmentation during training -- one could use the validation set to identify which augmentations improve the model's classification accuracy for each class, and then selectively augment the training data accordingly. We intend to explore this idea further in future work.

\section{Conclusion}
\label{sec:conclusion}
In this article we proposed a deep convolutional neural network architecture which, in combination with a set of audio data augmentations, produces state-of-the-art results for environmental sound classification. We showed that the improved performance stems from the combination of a deep, high-capacity model and an augmented training set: this combination outperformed both the proposed CNN without augmentation and a ``shallow'' dictionary learning model with augmentation. Finally, we examined the influence of each augmentation on the model's classification accuracy. We observed that the performance of the model for each sound class is influenced differently by each augmentation set, suggesting that the performance of the model could be improved further by applying class-conditional data augmentation.

\section*{Acknowledgment}
The authors would like to thank Brian McFee and Eric Humphrey for their valuable feedback, and Karol Piczak for providing details on the results reported in \cite{Piczak:EnvSoundCNN:MLSP:15}.
This work was partially supported by NSF award 1544753.

\ifCLASSOPTIONcaptionsoff
  \newpage
\fi

\bibliographystyle{IEEEtran}
\bibliography{IEEEabrv,salamon_cnn-aug-env_ieeespl_2016}

\end{document}